\documentstyle[preprint,aps]{revtex}
\begin{document}
\draft
\title{Why normal Fermi systems with sufficiently singular interactions
do not have a sharp Fermi surface}

\author{Peter Kopietz}
\address{
Institut f\H{u}r Theoretische Physik der Universit\H{a}t G\H{o}ttingen,\\
Bunsenstr.9, D-37073 G\H{o}ttingen, Germany}
\date{August 7, 1995}
\maketitle
\begin{abstract}
We use a bosonization approach to show that
the momentum distribution $n_{\bf{k}}$ of
normal Fermi systems with sufficiently singular interactions
is analytic in the vicinity of the non-interacting Fermi surface.
These include singular density-density interactions
that diverge in $d$ dimensions stronger than $ | {\bf{q}} |^{ - 2 ( d -1 ) }$
for vanishing momentum transfer ${\bf{q}}$, but also
fermions that are coupled to transverse gauge fields  in $ d < 3$.
\end{abstract}

\vspace{10mm}

\narrowtext

As first noticed by Bares and Wen\cite{Bares93}, singular
density-density interactions with Fourier transform $f_{\bf{q}} \propto |
{\bf{q}}|^{- \eta }$
destroy in $d$ dimensions the Fermi liquid state for $ \eta \geq 2 ( d-1 )$.
The case $\eta = 2 ( d-1 )$ is marginal and corresponds to a Luttinger liquid,
while
for $\eta > 2 ( d - 1 )$ one obtains normal metals which are neither Fermi
liquids nor
Luttinger liquids.
We shall call these metals {\it{strongly correlated quantum liquids}}.
The properties of these systems are not very well understood.
Certainly strongly correlated quantum liquids cannot be studied by means of
conventional many-body
perturbation theory, because the perturbative calculation of the
self-energy leads to power-law divergencies.
In the present work we shall use higher dimensional bosonization to calculate
the
momentum distribution $n_{\bf{k}}$ in these systems. We find that
for $ \eta > 2 ( d -1 )$ the momentum distribution
$n_{\bf{k}}$ does not have any singularities, so that a sharp Fermi surface
cannot be defined. We then argue that below three dimensions the
momentum distribution of electrons that are coupled to
transverse gauge fields has also this property.

Bosonization in arbitrary dimensions has recently
been discussed by a number of
authors\cite{Luther79,Haldane92,Houghton93,Castro94,Frohlich95,Kopietz94,Kopietz95}.
The fundamental geometric construction is the subdivision of the
Fermi surface into patches of area $\Lambda^{d-1}$.
With each patch one then associates a ``squat box'' $K^{\alpha}$
\cite{Houghton93} of radial hight
$\lambda \ll k_{F}$ and volume $\lambda \Lambda^{d-1}$, and partitions the
degrees of freedom
close to the Fermi surface into
the boxes $K^{\alpha}$. Here $k_{F}$ is
the Fermi wave-vector, and $\alpha$ labels the
boxes in some convenient ordering.
If the size of the patches is chosen small enough, then
the curvature of the Fermi surface can be {\it{locally}} neglected.
The essential motivation for this construction is
that it opens the way for a {\it{local linearization}} of the
non-interacting energy dispersion $\epsilon_{\bf{k}}$:
If ${\bf{k}}^{\alpha}$ points to the suitably defined center of box
$K^{\alpha}$, then for ${\bf{k}} \in K^{\alpha}$ we may shift
${\bf{k}} = {\bf{k}}^{\alpha} + {\bf{q}}$, and expand
$\epsilon_{{\bf{k}}^{\alpha} + {\bf{q}} } - \mu \approx {\bf{v}}^{\alpha} \cdot
{\bf{q}}$,
where $\mu$ is the chemical potential and ${\bf{v}}^{\alpha}$ is the local
Fermi velocity.
At high densities and for interactions that are dominated by
small momentum transfers
the linearization implies {\it{in arbitrary dimension}}
{\it{a large-scale cancellation between
self-energy and vertex-corrections}} (generalized closed loop
theorem\cite{Kopietz95}),
so that the entire perturbation series
can be summed in a controlled way.
The final result for the Matsubara-Green's function
$G ( {\bf{k}} , i \tilde{\omega}_{n} )$ of the
interacting many-body system is then
 \begin{eqnarray}
 G ( {\bf{k}} , i \tilde{\omega}_{n} )  & = &  \sum_{\alpha} \Theta^{\alpha} (
{\bf{k}} )
 \int d {\bf{r}} \int_{0}^{\beta} d \tau
 e^{ - i [ ( {\bf{k}} - {\bf{k}}^{\alpha} ) \cdot  {\bf{r}}
 - \tilde{\omega}_{n}  \tau  ] }
 \nonumber
 \\
 & \times &
 {{G}}^{\alpha}_{0} ( {\bf{r}}  , \tau  )
 e^{ Q^{\alpha} ( {\bf{r}} , \tau )
  }
 \; \; \; ,
 \label{eq:Gkres2}
 \end{eqnarray}
where $\Theta^{\alpha} ( {\bf{k}}  )$ is unity of ${\bf{k}} \in K^{\alpha}$ and
vanishes
otherwise, and
 \begin{equation}
 G^{\alpha}_{0} ( {\bf{r}}  , \tau  )
 =
 \delta^{(d-1)}_{\Lambda} ( {\bf{r}}_{\bot}  )
 \left( \frac{ - i}{2 \pi} \right)
 \frac{1}
 {
  r_{\|}
 + i | {\bf{v}}^{\alpha} |  \tau }
 \; \; \; ,
 \label{eq:Gpatchreal1}
 \end{equation}
 \begin{equation}
 Q^{\alpha} ( {\bf{r}} , \tau )  =
 \frac{1}{\beta {{V}}} \sum_{ q }  f^{RPA}_{q}
  \frac{ 1 -
  \cos ( {\bf{q}} \cdot  {\bf{r}}
  - {\omega}_{m}  \tau  )
 }
 {
 ( i \omega_{m} - {\bf{v}}^{\alpha} \cdot {\bf{q}} )^{2 }}
 \label{eq:DWph}
 \; \; \; .
 \end{equation}
Here $\beta$ is the inverse temperature, $V$ is the volume of the system,
$q = [ {\bf{q}} , i {\omega}_{m} ]$,
and $r_{\|} = {\bf{r}} \cdot \hat{\bf{v}}^{\alpha}$, with
$\hat{\bf{v}}^{\alpha} = {\bf{v}}^{\alpha} / | {\bf{v}}^{\alpha} |$.
The fermionic Matsubara frequencies are denoted by
$\tilde{\omega}_{n} = 2 \pi [ n + \frac{1}{2} ] / \beta$, and the bosonic ones
are
$\omega_m = 2 \pi m / \beta$.
For length scales $| {\bf{r}}_{\bot} | \gg \Lambda^{-1}$ the
function $\delta^{(d-1)}_{\Lambda} ( {\bf{r}}_{\bot} )$
can be treated as a $d-1$-dimensional Dirac-$\delta$ function of the components
of $ {\bf{r}} $
orthogonal to $\hat{\bf{v}}^{\alpha}$.
The effect of the interactions is contained in
$Q^{\alpha} ( {\bf{r}} , \tau )$, which depends exclusively
on the random-phase approximation $f^{RPA}_{q}$
for the interaction.
In the functional-integral formulation of
bosonization\cite{Frohlich95,Kopietz94,Kopietz95,Fogedby76},
$Q^{\alpha} ( {\bf{r}} , \tau )$ can be interpreted simply as the usual
{\it{Debye-Waller factor}}
that arises in a Gaussian average.
A result similar to Eqs.\ref{eq:Gkres2}-\ref{eq:DWph} has
also been obtained by Castellani, Di Castro and Metzner\cite{Castellani94a}
by means of a non-perturbative approach based on Ward
identities\cite{Metzner93}.

We consider density-density interactions of the form
$f_{\bf{q}} = f_{0} ( q_{c} / | {\bf{q}|} )^{\eta} e^{- |{\bf{q}}| / q_{c} }$,
with
$\eta > 0$ and $q_{c} \ll k_{F}$.
After lengthy but straight-forward algebra we obtain from Eq.\ref{eq:DWph}
for the leading asymptotic behavior of the equal-time
Debye-Waller factor at large distances
 \begin{equation}
 Q^{\alpha} ( r_{\|} \hat{\bf{v}}^{\alpha} , 0 )
 \sim
 \left\{
 \begin{array}{ll}
  R_{d, \eta }
  & \mbox{  for $ \eta <   2 (d-1) $}
 \; \; \; ,
 \\
 - \gamma_{d} \ln ( q_{c} | r_{\|} | )
 &  \mbox{ for $ \eta = 2 ( d-1 )$}
 \; \; \; ,
 \\
- \beta_{d , \eta }  ( q_{c} | r_{\|}| )^{  \frac{\eta}{2} - d + 1 }
&  \mbox{ for $ \eta > 2 ( d-1 )$}
\; \; \; ,
\end{array}
\right.
\label{eq:Qasymsummary}
\end{equation}
where we have assumed for simplicity that the Fermi surface
is spherically symmetric.
Here $R_{d , \eta }$, $\gamma_{d}$ and $\beta_{d , \eta }$ are  finite real
numbers
that depend not only on $d$ and $\eta$, but also in $F_{0} \equiv \nu f_{0}$,
where $\nu$ is the $d$-dimensional density of states at the Fermi energy.
Explicit expressions for $R_{d , \eta}$,
$\gamma_{d}$ and $\beta_{d , \eta }$ can be written down in arbitrary $d$,
but are  omitted here for brevity, because the precise values
for these quantities is not essential in this work.
We would like to point out however, that
in arbitrary dimension there exists
a critical  value $\eta_{d}^{\ast} > 2 ( d-1 )$
where $\beta_{d , \eta^{\ast}_{d}}$ diverges.
It is not clear if this divergence signals a physical
instability of the metallic state, or simply arises from the
inadequacy of the bosonization approach.
For example, in $d=1$ we obtain for $\eta > 0$
 \begin{eqnarray}
 \beta_{1, \eta}   & = &
 \frac{ \sqrt{F_{0}} }{2} \int_{0}^{\infty} dx \frac{ 1 - \cos x}{ x^{1 + \eta
/2 }}
 \nonumber
 \\
 & = &
 \frac{\sqrt{F_{0}}}{\eta} {  \Gamma ( 1 - \frac{\eta }{ 2} )
 \cos \left( \frac{\pi \eta }{4} \right)}
 \; \; \; .
 \end{eqnarray}
Because $\beta_{1, \eta} \rightarrow \infty$ for
$\eta \rightarrow 4 $, the
static Debye-Waller factor
is divergent for $\eta \geq 4$,
so that $\eta_{1}^{\ast} = 4$.
More generally, for singular interactions in arbitrary $d$
it is easy to show by simple power counting that
$\eta_{d}^{\ast} = 2 ( d + 1)$.
However,  the finiteness of the static Debye-Waller factor
does {\it{not}} imply that
for $\tau \neq 0$ the function $Q^{\alpha} ( r_{\|} \hat{\bf{v}}^{\alpha} ,
\tau  )$
remains finite as well.  For example,
in $d=1$ the standard bosonization procedure\cite{Tomonaga50} leads to an
integral over the term
 \begin{equation}
 \frac{ \cos ( q_{\|} r_{\|} )}{q_{\|}^{1 + \eta /2} }
 \left[ 1 - e^{ - \sqrt{F_{0}} q_{c}^{\eta /2} v_{F} | \tau | q_{\|}^{1 - \eta
/2 }} \right]
 \sim
 \frac{ \sqrt{F_{0}} q_{c}^{\eta /2} v_{F} | \tau |  }{ q_{\|}^{\eta } }
 \; \; \; ,
 \end{equation}
where the expansion of the exponential for small $q_{\|}$ is justified as long
as $\eta < 2$.
Obviously, the singularity at small $q_{\|}$ is not integrable for $\eta  \geq
1$,
so that in $d = 1$ the full Green's function
can only be calculated via bosonization for $\eta <  1$.
In the rest of this paper we shall assume that $0 < \eta < \eta_{d}^{\ast} = 2
( d+1 )$, so that
the bosonization result for the equal-time Green's function
is free of divergencies.

Given the equal time Green's function, we may calculate the momentum
distribution
$n_{\bf{k}} = \frac{1}{\beta} \sum_{n} G ( {\bf{k}} , i \tilde{\omega}_{n} )$
in the
vicinity of the Fermi surface.
Using Eq.\ref{eq:Gkres2}
and shifting ${\bf{k}} = {\bf{k}}^{\alpha} + {\bf{q}}$, it is easy to show that
for $ | {\bf{q}} | \ll q_{c}$
\begin{equation}
 \delta n^{\alpha}_{ {\bf{q}} }  \equiv
 n_{ \bf{k}^{\alpha} - {\bf{q}} } -
 n_{ \bf{k}^{\alpha} + {\bf{q}} } =
  \frac{2}{\pi}
  \int_{ 0 }^{\infty}
 d {r_{\|}} \frac{ \sin ( q_{\|} r_{\|}  ) }{ r_{\|} }
 e^{ Q^{\alpha} ( r_{\|} \hat{\bf{v}}^{\alpha} , 0 ) }
 \; \; \; .
 \label{eq:occup4}
 \end{equation}
Note that $\delta n^{\alpha}_{\bf{q}}$
depends only on the projection $q_{\|}$ of ${\bf{q}}$
that is normal to the Fermi surface,  because
this component corresponds to a crossing of the Fermi surface
and can therefore be associated with a
possible discontinuity.
Because Eq.\ref{eq:Qasymsummary} is valid
for  $r_{\|}
{ \raisebox{-0.5ex}{$\; \stackrel{>}{\sim} \;$}}
q_{c}^{-1}$, we separate from
Eq.\ref{eq:occup4} the non-universal short-distance regime and obtain
for $|q_{\|} | \ll q_{c}$
after rescaling the integration variables
 \begin{equation}
 \delta n^{\alpha}_{ {\bf{q}} }  =
 z^{\alpha} \frac{ q_{\|} }{q_{c} }
  +
  \frac{2}{  \pi}
  \int_{   1 }^{\infty}
 d {x} \frac{ \sin   (  \frac{q_{\|}}{q_{c}}   x )   }{ x }
 e^{ Q^{\alpha} (    \frac{x}{ q_{c}}   \hat{\bf{v}}^{\alpha}  , 0 ) }
 \; \; \; ,
 \label{eq:occup4c}
 \end{equation}
where the non-universal constant $z^{\alpha}$ is given by
 \begin{equation}
 z^{\alpha} =
  \frac{2}{\pi }
  \int_{ 0 }^{ 1 }
 d {x}
 e^{ Q^{\alpha} ( \frac{x}{q_{c}} \hat{\bf{v}}^{\alpha} , 0 ) }
 \label{eq:Calphanudef}
 \; \; \; .
 \end{equation}
For $| q_{\|} | \ll q_{c} $ we may
substitute
in the second term of Eq.\ref{eq:occup4c}
the asymptotic
expansion of ${ Q^{\alpha} ( \frac{x}{q_{c}} \hat{\bf{v}}^{\alpha}  , 0 ) }$
for large
$\frac{x }{  q_{c}}$,
see Eq.\ref{eq:Qasymsummary}.
For $ \eta < 2 ( d-1 )$ we obtain
 \begin{equation}
 \delta n^{\alpha}_{ {\bf{q}} }  = e^{R^{\alpha}} sign (q_{\|})
  \; \; \; ,  \; \; \mbox{ $\eta < 2 ( d-1)$ }
  \label{eq:deltanfinal1}
  \; \; \; .
  \end{equation}
This is the usual Fermi  liquid behavior: the discontinuity of the
momentum distribution at point ${\bf{k}}^{\alpha}$
is given by the quasi-particle residue $Z^{\alpha} = e^{R^{\alpha}}$.
Because $R^{\alpha} $ is negative\cite{Kopietz94,Kopietz95}, we have
$0 < Z^{\alpha} < 1$.
Note that for small $q_{\|}$ the first term in Eq.\ref{eq:occup4c} is
negligible.
In the marginal case $ \eta = 2 ( d-1 )$ we obtain
for $| q_{\|} | \ll q_{c} $
 \begin{equation}
 \delta n^{\alpha}_{ {\bf{q}} }   \sim
 z^{\alpha} \frac{ q_{\|} }{ q_{c}}
 +
  \frac{ 2 sign ( q_{\|} )  }{\pi}
  \left| \frac{q_{\|} }{ q_{c} } \right|^{\gamma_{d} }
  \int_{  |q_{\|}|/ q_{c}  }^{\infty}
 d {x} \frac{ \sin x }{ x^{1+ \gamma_{d} } }
 \label{eq:occup7}
 \; \; \; .
 \end{equation}
{}From this expression it is easy to show that
to leading order
 \begin{equation}
 \delta n^{\alpha}_{ {\bf{q}} }  \sim
 \left\{
 \begin{array}{ll}
  sign ( q_{\|} )  | \frac{ q_{\|}  }{  q_{c} } |^{\gamma_{d}}
 & \mbox{  for $\gamma_{d} \ll 1$ }
 \\
 \frac{2}{\pi} ( \frac{q_{\|}}{ q_{c} } )
 \ln ( \frac{ q_{c}}{|q_{\|}| } )
 & \mbox{ for $\gamma_{d} = 1$ }
 \\
  ( z^{\alpha}
  + \frac{2}{ \pi ( \gamma_{d} - 1) }  )
  \frac{ q_{\|}}{q_{c} }
 & \mbox{  for $\gamma_{d} > 1$ }
 \end{array}
 \right.
 \; \; \; , \; \; \; \eta = 2 ( d-1 ) \; \; \; .
 \end{equation}
An algebraic singularity in the  momentum distribution
is characteristic for Luttinger liquids\cite{Tomonaga50}.
Note, however,  that for $\gamma_{d} = 1$
the singularity is only logarithmic, and that for
$ \gamma_{d} > 1$ the leading term is even analytic.
Although in this case there are non-analyticities in the higher order
corrections,
one may wonder whether for $\gamma_{d} > 1$ the bosonization approach is
consistent.
Recall that
we have linearized the energy dispersion
``at the Fermi surface'',
thus implicitly assuming that the Fermi surface can somehow be defined.
In the case of a Fermi liquid the finite
discontinuity of the momentum distribution leads to a unique definition of the
Fermi surface.
For a  Luttinger liquid one may define the Fermi surface
as the set of points in momentum space where
$ n_{\bf{k}}$ has an algebraic singularity.
However, for $\gamma_{d} > 1$
it is at least not quite satisfactory that one has
to rely on asymptotically irrelevant
corrections to define the Fermi surface.
This point becomes even more questionable if we consider
the case $ \eta > 2 ( d-1 ) $. Then we obtain
from Eqs.\ref{eq:occup4c} and \ref{eq:Qasymsummary}
 \begin{equation}
 \delta n^{\alpha}_{\bf{q}} \sim
 z^{\alpha} \frac{ q_{\|}}{q_{c}}
 +
 \frac{2}{\pi}
 \int_{ 1 }^{\infty} d x \frac{ \sin ( \frac{q_{\|}}{ q_{c} } x)  }{ x }
 \exp \left[  - \beta_{d, \eta } x^{ \frac{\eta}{2} - d + 1  } \right]
 \; \; \; .
 \label{eq:discontEX1}
 \end{equation}
The crucial observation is now that
{\it{the stretched exponential vanishes faster than
any power}}, so that the integral can be done by expanding
$\sin ( \frac{ q_{\|} }{q_{c}} x )$
under the integral sign and exchanging the order of
integration and summation.
It immediately follows that
 $\delta n^{\alpha}_{\bf{q}} $ is for $\eta > 2 ( d-1)$ {\it{an analytic
function
of}} $q_{\|}$. To leading order we have
 \begin{equation}
 \delta n^{\alpha}_{\bf{q}}
 =
 \left[ z^{\alpha}
 + \frac{2}{\pi} \int_{ 1 }^{\infty} dx
 \exp [ - \beta_{d , \eta }  x^{ \frac{\eta}{2} -d + 1} ] \right]
 \frac{ q_{\|}}{q_{c}}
 +
 O ( q_{\|}^2 )
 \; \; \; .
 \label{eq:discontEX2}
 \end{equation}
If $\beta_{d , \eta }$ or $ \frac{\eta}{2} - d + 1$ is small, the second term
is dominant,
because then the integral is determined
by the large-$x$ regime. Then we may extend
the lower limit to zero and obtain
 \begin{equation}
 \delta n^{\alpha}_{\bf{q}} =
 \frac{4}{\pi}
 \frac{   \Gamma ( \frac{2}{ \eta - 2(d - 1) } )
 \beta_{d, \eta }^{ - \frac{2}{ \eta - 2(  d - 1)  } }
 }{
  \eta - 2( d - 1)
 }
  \frac{q_{\|} }{q_{c}}
 +
 O ( q_{\|}^2 )
 \; \; \; ,
 \; \; \; \eta > 2 ( d-1 )
 \; \; \; .
 \label{eq:discontEX3}
 \end{equation}
Hence, there is no singularity whatsoever in the momentum distribution,
so that a sharp Fermi surface simply cannot be defined.
The complete destruction of the Fermi surface in strongly correlated Fermi
systems
is certainly not a special feature of the singular interactions
studied in the present work.
For example, models with correlated hopping\cite{Hirsch89,Bariev93}
show similar behavior.

The obvious question is now if for $ \eta > 2 ( d-1 )$the bosonization approach
is consistent or not.
Before addressing this question, let us briefly recall that
in disordered systems the situation is
precisely the same\cite{Abrikosov63}.
Also here the average momentum distribution in the vicinity
of the Fermi surface of the clean system does not have any singularities.
In this case the thickness of the shell where $n_{\bf{k}}$ drops from unity
to zero is given by the inverse mean free patch $\ell^{-1}$.
Because in good metals $\ell^{-1} \ll k_{F}$, the
``Fermi surface'' is defined in the sense that
outside the limits of a thin shell in momentum space
the derivative of $n_{\bf{k}}$ is negligibly small.
Because in the laboratory  impurities can never be completely
eliminated, this definition of the Fermi surface does justice
to the experimental reality, although
it is impossible to define a surface
in the strict mathematical sense.
Clearly, the ``Fermi surface'' in the present problem
should be defined analogously:
as long as the thickness $k_{S}$ of the shell
where the momentum distribution varies appreciably is small compared with
$k_{F}$, it is meaningful to talk about a
{\it{smeared Fermi surface}}, or, more accurately,
a {\it{Fermi shell}}.
The condition $k_{S} \ll k_{F}$
is sufficient to make the bosonization approach internally consistent, because
then it does not matter
at which location within the Fermi shell
the non-interacting energy dispersion has been linearized.
This point of view has also been emphasized in the classic
paper by Tomonaga\cite{Tomonaga50}.

The properties of strongly correlated  quantum liquids  are not
only of academic interest. Physical manifestations of
such an unusual metallic state might have been observed in
the normal-state of the cuprate superconductors\cite{Ioffe89}, or in
half filled quantum Hall systems\cite{Halperin93}.
Recent theoretical models for these systems
involve the coupling between
electrons and transverse gauge fields, which at the fermionic level
leads to an effective current-current interaction.
In the perturbative calculation of the fermionic self-energy
this rather singular interaction gives rise to divergencies
which are stronger than logarithmic.
The singular nature of the current-current interaction
is essentially a consequence of gauge invariance, which implies
that, in the absence of superconducting instabilities, the
gauge field field cannot be screened in the static
limit\cite{Holstein73,Pethick89}.
Recently this problem and its generalizations
to arbitrary dimension $d$
has been re-examined by a number of authors
\cite{Khveshchenko93,Gan93,Plochinski94,Nayak94,Ioffe94,Castellani94,Kwon94,Chakravarty95,Kopietz95c}.
Although the applicability of the higher dimensional bosonization
approach to this problem has been
questioned\cite{Ioffe94,Castellani94},
there exist other independent non-perturbative calculations which
confirm the bosonization result\cite{Khveshchenko93}.
It is perhaps fair to say that at present the issue is
far from being settled.
For electrons coupled to the Maxwell-field the higher-dimensional
bosonization approach\cite{Kwon94,Kopietz95c} implies that $d=3$
is a marginal dimension in the problem, in agreement with
the renormalization group calculations\cite{Gan93,Chakravarty95}.
For $d < 3$ bosonization predicts
$Q^{\alpha} ( r_{\|} \hat{\bf{v}}^{\alpha} , 0 ) \propto
- ( \kappa_{d} r_{\|} )^{ \frac{3-d}{3} }$, where the momentum scale
$\kappa_{d}$
is given in Ref.\cite{Kopietz95c}.
{}From our analysis given above it is clear that
the stretched  exponential behavior implies that below three dimensions the
coupling to the transverse gauge field completely washes out the Fermi surface.

In summary, we have shown that in strongly correlated quantum liquids
the momentum distribution is an analytic function
close to the non-interacting Fermi surface.
In these systems the concept of a Fermi surface
must be replaced by a Fermi shell.
The perhaps most important physical realization of such a system
are half filled quantum Hall systems.
Although recent experiments suggest that in these systems there is a
well-defined
Fermi surface\cite{Willet93}, the finite smearing scale $k_{S}$ might be beyond
experimental
resolution.  We have also argued that,
at least for not too singular interactions,
the bosonization approach remains consistent
as long as the  thickness $k_{S}$ of the Fermi shell is small
compared with $k_{F}$.

I would like to thank Cornelia Buhrke, Guillermo Castilla, and Jean Lapointe
for  enjoying with me the 1995 APS March Meeting in San Jose, where this work
was conceived.
I am also grateful to K. Sch\H{o}nhammer for many discussions and
fruitful collaborations.

%

\end{document}